\documentclass[12pt,a4paper]{article}%

\usepackage{amsmath,amssymb,amsfonts}
\usepackage{caption2}
\usepackage[]{hyperref}
\usepackage[pdftex]{color,graphicx}

\numberwithin{equation}{section} 
\setcaptionmargin{1cm}

\begin{document}
\begin{titlepage}
\vskip 1.0cm
\begin{center}
{\Large \bf A Non Standard Supersymmetric Spectrum} \vskip 1.0cm {\large Riccardo
Barbieri$^a$, Enrico Bertuzzo$^a$, Marco Farina$^a$, Paolo Lodone$^a$\ \   and Duccio Pappadopulo$^b$} \\[1cm]
{\it $^a$ Scuola Normale Superiore and INFN, Piazza dei Cavalieri 7, 56126 Pisa, Italy} \\[5mm]
{\it $^b$ Institut de Th\'eorie des Ph\'enom\`enes Physiques, EPFL,  CH--1015 Lausanne, Switzerland}\vskip 1.0cm
\end{center}
\begin{abstract}
Taking a bottom-up point of view and focussing on the lack of signals so far in the Higgs and in the flavour sectors, we argue in favour of giving consideration to  supersymmetric extensions of the Standard Model where the lightest Higgs boson has a mass between 200 and 300 GeV and the first two generations of s-fermions are above 20 TeV.  After examining the simplest extensions of the Minimal Supersymmetric Standard Model that allow this in a natural way, we summarize the main consequences of this pattern of masses at the LHC and we analyze the consequences of a heavier than normal Higgs boson for Dark Matter.
\end{abstract}
\end{titlepage}


\section{Motivations and general programme}
\label{int}

Phenomenological supersymmetry allows to incorporate the perturbative Standard Model in a  theory which solves the hierarchy problem all the way up to the Planck scale, with a  potentially  successful  description of gauge coupling unification. The consistency with the ElectroWeak Precision Tests (EWPT) of the Standard Model (SM) with a relatively light Higgs boson adds support to this view, making  the test of the Minimal Supersymmetric Standard Model (MSSM) a crucial task of the LHC. 
This is a meaningful straight path to be followed for particle physics in the next few years. While remaining in the context of phenomenological supersymmetry, however, the lack of signals so far 
both in the Higgs and in the flavour sectors have raised and continue to raise questions.

That the Higgs problem of the MSSM be a naturalness problem is too well known to be recalled here in detail: the sensitivity of the Fermi scale, as determined by the Higgs potential of the MSSM, to the average stop masses makes it unnatural to raise the mass of the lightest scalar, $h$, much above the tree level bound, $m_h \leq m_Z |\cos{2\beta}|$, in potential conflict with the LEP bounds. Is the flavour problem a naturalness problem as well?  Given the little we understand about flavour, this is not the easiest question to answer. Let us take the view, however, as put forward by many authors in the nineties \cite{Dine:1990jd} - \cite{Barbieri:1997tu}, that the supersymmetric flavour problem may have something to do with  a hierarchical structure of  s-fermion masses:  the first two generations significantly heavier than the third one. How much heavier can now become a naturalness problem, depending on the bounds that the sfermion masses have to satisfy \cite{Barbieri:1987fn}\cite{Dimopoulos:1995mi}.  Here we argue about the possibility that the two issues, ``the Higgs problem" and ``the flavour problem", be related naturalness problems, that may be addressed at the same time by properly extending the MSSM.

\begin{figure}[tb]
\centering
\includegraphics[width=10cm]{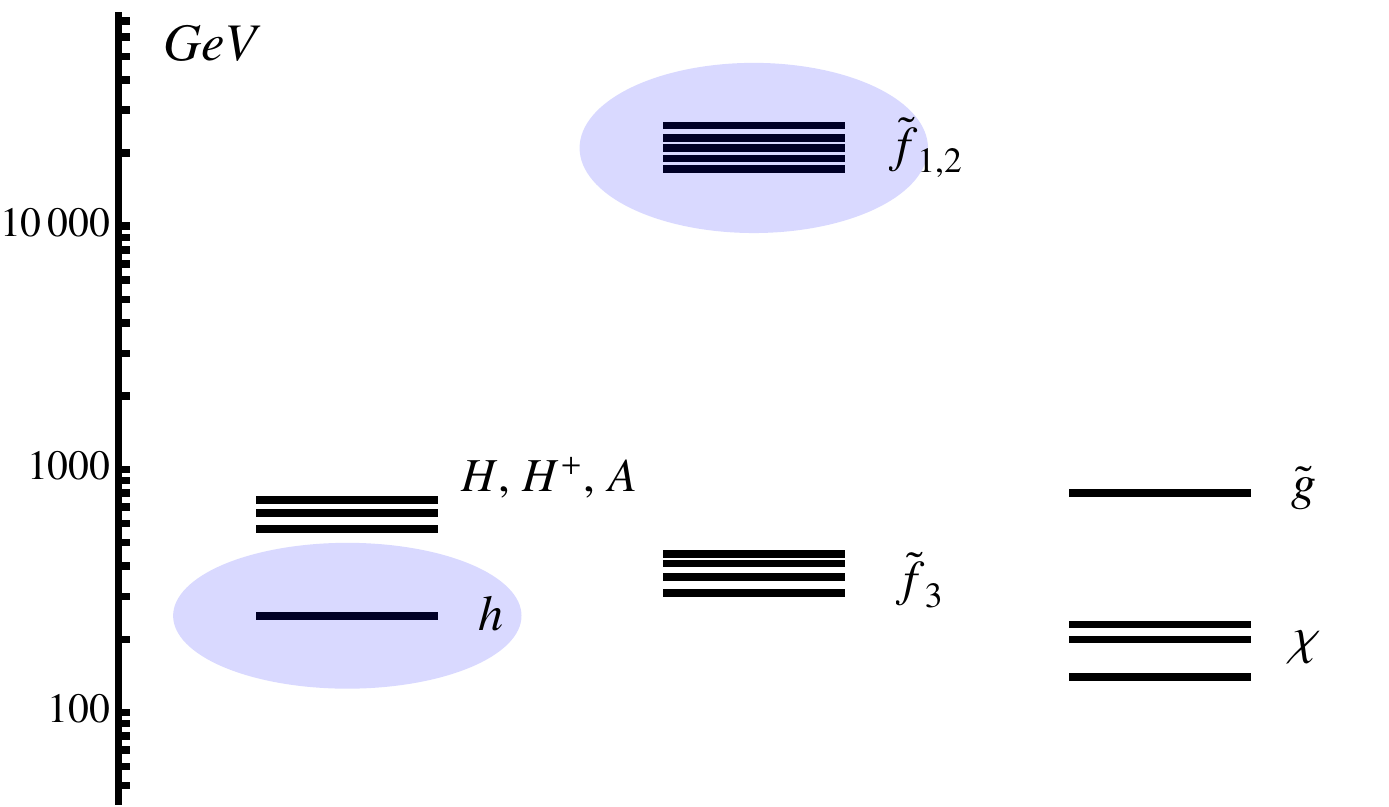} \caption{{\small A representative Non Standard Supersymmetric Spectrum with $m_h = 200\div 300$ GeV and $m_{\tilde{f}_{1,2}}\gtrsim 20$ TeV.}}
\label{spettro2}
\end{figure}

Let us insist on the supersymmetric flavour problem in connection with a hierarchical s-fermion spectrum. As well known, without degeneracy nor alignment between the first two generations of squarks, $m_{\tilde{q}_{1,2}}$, the consistency with the $\Delta S = 2$ transitions, 
both real and especially imaginary, would require values of $m_{\tilde{q}_{1,2}}$ far too big to be natural. 
Relatively mild assumptions on all the s-fermion masses of the first two generations, on the other hand, as recalled later, allow to satisfy the various flavour constraints by smaller values of $m_{\tilde{f}_{1,2}}$  that may be considered if they are natural or not, hence the potential connection with the Higgs mass problem. In formulae, the two naturalness constraints ($1/\Delta$ is the amount of fine tuning as defined in the usual way \cite{Barbieri:1987fn}, $m_{\tilde{t}}$ is the average stop mass):
\begin{equation}
\frac{m_{\tilde{t}}^2}{m_h^2}
\frac{\partial m_h^2}{\partial m_{\tilde{t}}^2} < \Delta
\label{natbounds1}
\end{equation}
\begin{equation}
\frac{m^2_{\tilde{f}_{1,2}}}{m_h^2}
\frac{\partial m_h^2}{\partial m^2_{\tilde{f}_{1,2}}} < \Delta
\label{natbounds2}
\end{equation}
must be considered together and the corresponding bounds might be reduced to an acceptable level by pushing up the theoretical value of $m_h$, on which the level of fine tuning depends at least quadratically\footnote{Note that replacing the physical Higgs mass $m_h$ with the $Z$ mass or with any of the soft mass parameters for the Higgs doublets does not change the naturalness constraints on $m_{\tilde{t}}$ or on $m_{\tilde{f}_{1,2}}$, at least as long as the other physical Higgs bosons are not too close in mass to the lightest one, $h$, as we consider in the following for good phenomenological reasons. On this, see e.g. \cite{Barbieri:2006bg}.}.
 Ways to push up $m_h$ even by a significant amount, between 200 and 300 GeV, have already been put forward \cite{Espinosa:1998re} - \cite{Barbieri:2006bg}.  Whether and how the flavour problem can also be attacked in this manner  is a model dependent question that we are going to analyze in various cases proposed in the literature. In summary, and  as an anticipation, we seek for models where a typical spectrum like the one shown in Fig.  \ref{spettro2}  can be naturally implemented. 

\section{Hierarchical s-fermion masses and flavour physics: a summary}
\label{Hier}

A way to summarize the potential connection between the supersymmetric flavour problem and hierarchical s-fermion masses is the following\footnote{For a recent analysis see \cite{Giudice:2008uk}. Notice however that in that paper  one always considers $\delta_{LL} >> \delta_{RR}$ or viceversa.}.
\begin{itemize}
\item
Without degeneracy nor alignment the bounds that the first two generations of squark masses would have to satisfy to be compatible with the flavour constraints, mostly from $\Delta S =2$ transitions, are in the hundreds of TeV, with weak dependence on the much lighter gaugino masses. 
On the other hand, if we assume degeneracy and alignment of order of the Cabibbo angle, i.e. in terms of the standard notation:
\begin{equation}
\delta^{LL}_{12} \approx \frac{|m^2_1 - m^2_2|}{(m^2_1 + m^2_2)/2} \approx \lambda \approx 0.22,
\label{condit}
\end{equation}
and $\delta^{LL} \approx  \delta^{RR} >> \delta^{LR}$,
then the bounds are significantly reduced to:
\begin{equation}
Real~\Delta S = 2 \Rightarrow  m_{\tilde{q}_{1,2}} \gtrsim 18~TeV   \, ,
\end{equation}
\begin{equation}
Im~\Delta S = 2,~\sin{\phi_{CP}}\approx 0.3 \Rightarrow  m_{\tilde{q}_{1,2}} \gtrsim 120~TeV  \, .
\end{equation}
Furthermore if
$\delta^{LL} >> \delta^{RR} , \delta^{LR}$
(or $\delta^{RR} >> \delta^{LL} , \delta^{LR}$), these bounds are replaced in the strongest cases by:
\begin{equation}
\Delta C = 2 \Rightarrow  m_{\tilde{q}_{1,2}} \gtrsim 3~TeV  
\label{3TeV}
\end{equation}
\begin{equation}
Im~\Delta S = 2,~\sin{\phi_{CP}} \approx 0.3 \Rightarrow  m_{\tilde{q}_{1,2}} \gtrsim 12~TeV 
\label{12TeV} 
\end{equation}
 from CP conserving or CP violating effects respectively.

\item
The exchange of the third generation of s-fermions may also produce too big flavour effects unless the off-diagonal $\delta_{i3}, i=1,2$ are small enough. If for example we assume a correlation between the off-diagonal elements and the ratio of the diagonal masses of the type:
\begin{equation}
\delta^{LL}_{i3} \approx \frac{m^2_{\tilde{f}_3}}{m^2_{\tilde{f}_i}},
\end{equation}
a dominant constraint comes from $B-\overline{B}$ mixing:
\begin{equation}
\Delta B = 2 \Rightarrow  m_{\tilde{q}_{1,2}} \gtrsim  6~TeV (\frac{m_{\tilde{q}_{3}}}{500~GeV})^{1/2}.
\end{equation}
\end{itemize}
Similar or weaker constraints are obtained from the Electric Dipole Moments.

\begin{figure}[hbt]
\begin{center}
\begin{tabular}{cc}
\includegraphics[width=0.44\textwidth]{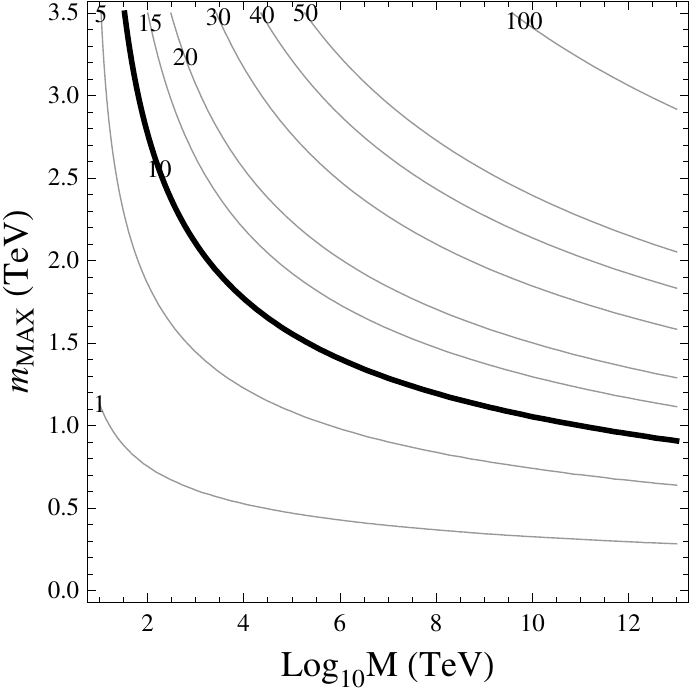} &
\includegraphics[width=0.44\textwidth]{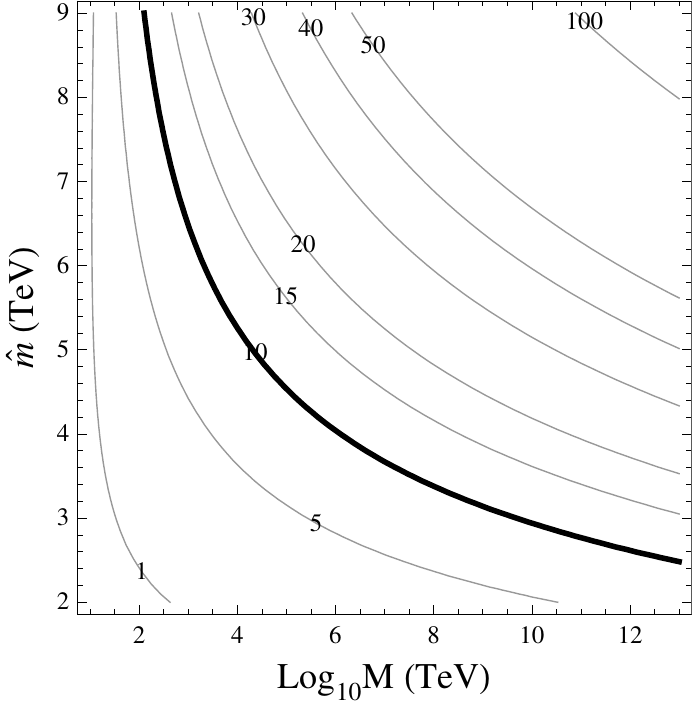}
\end{tabular}
\end{center}
\caption{{\small Upper bounds for different $\Delta = 1, \dots, 100$ on the masses of the first and second generation scalars as function of the scale $M$ at which they are generated. Left: no special condition at $M$. Right:  degenerate masses at $M$, at least within $SU(5)$ multiplets.}}
\label{naturalnessMSSM}
\end{figure}

As said,   too little is known about flavour to be able to draw any firm conclusion.  Yet the pattern of charged fermion masses makes it conceivable that  approximate flavour symmetries be operative to justify some of the assumptions made above and therefore the corresponding bounds. In turn, at least as an orientation, it is useful to compare them with the naturalness constraints that limit the sfermion masses from above \cite{Dimopoulos:1995mi}\cite{Giudice:2008uk}. In the MSSM case, this is shown in Fig.s \ref{naturalnessMSSM} as function of the scale $M$ at which the soft masses are generated. In the figure on the left  the bound is on the heaviest among the sfermion masses of the first two generations, when the source of the renormalization of $m_h$, relevant to (\ref{natbounds2}),
 is a one-loop induced hypercharge Fayet-Iliopouolos term:
\begin{equation}
Tr(Y\tilde{m}^2) = Tr(\tilde{m}^2_Q + \tilde{m}^2_D -2 \tilde{m}^2_U -\tilde{m}^2_L +\tilde{m}^2_E)
\end{equation}
without particular initial conditions on the individual terms. When the Fayet-Iliopouplos term vanishes, then the dominant effect on $m_h$ comes from two loops. In the figure on the right side we show the bound on the (approximately degenerate) sfermion masses of the first two generations assuming them to be degenerate, at least within $SU(5)$ multiplets, at the scale $M$ where the renormalization group flow starts.

All this shows that in the MSSM, without giving up naturalness, the flavour problem can perhaps  be addressed by a hierarchical structure of the sfermion masses only if rather specific assumptions about their flavour structure are made, definitely stronger than the ones described above. While this is not excluded, we find it useful the reconsider the same problem in a broader context than the MSSM.

\section{Supersymmetry without a light Higgs boson}
\label{nolightH}
\subsection{Cases of interest}

Extensions of the MSSM have been studied that allow a significant increase of the mass of the lightest Higgs scalar, say above 200 GeV. This goes from the consideration of the MSSM as an effective Lagrangian with the inclusion of supersymmetric non-renormalizable operators \cite{Polonsky:2000rs}\cite{Casas:2003jx}\cite{Brignole:2003cm} to the design of specific models, valid up to a large scale, that try to 
keep the success of perturbative gauge coupling unfication. Here we take an intermediate view. On one side we want to keep manifest  consistency with the EWPT, which we do by requiring a minimum value of the scale $\Lambda$ at which perturbativity holds at least up to $5-10$ TeV. In particular this leads us not to consider raising significantly the Higgs boson mass by the inclusion of higher dimensional operators.
On the other side, in line with a typical bottom-up viewpoint, we do not seek for a complete description of the physics all the way up to (possible) unification. A representative of some of the attempts that satisfy these criteria is the following\footnote{For details on a recent comparative study on the models relevant to this entire Section see \cite{Lodone:2010kt}.}:
\begin{itemize}
\item {\it Extra $U(1)$ factor.} \cite{Batra:2003nj} The MSSM is extended to include an extra $U(1)$ factor with coupling $g_x$ and charge $\pm 1/2$ of the two standard Higgs doublets.  The extra gauge factor, under which also the standard matter fields are necessarily charged,  is broken by the vevs of a pair of extra scalars, $\phi$ and $\phi_c$, each in one chiral extra singlet, at a significantly higher scale than $v$. The upper bound on the mass of the lightest Higgs scalar now becomes:
\begin{equation}
m_h^2 \leq (m_Z^2 +\frac{g_x^2 v^2}{2(1+\frac{M_X^2}{2 M_\phi^2})})\cos^2{2\beta}
\label{mhU1}
\end{equation}
where $M_X$ is the mass of the new gauge boson and $M_\phi$ is the soft breaking mass of the scalars $\phi$, or $\phi_c$, taken approximately degenerate.
\item {\it Extra $SU(2)$ factor.} \cite{Batra:2004vc}\cite{Maloney:2004rc} In this case the standard ElectroWeak gauge group is  extended to $SU(2)_I\times SU(2)_{II}\times U(1)_Y$ with $SU(2)$ couplings $g_I$ and $g_{II}$.
For simplicity we take that all the standard matter fields, and so the two Higgs doublets, only transform under one of the $SU(2)$-factors  (but will comment later on on this property). The two $SU(2)$ are broken down,  at a scale about two orders of magnitude higher than $v$,  to the diagonal $SU(2)$ subgroup by the vev of a chiral multiplet $\Sigma$ transforming as $(2,2)$. In such a case the upper bound on the Higgs mass becomes:
\begin{equation}
m_h^2 \leq  m_Z^2 \frac{g^{\prime 2} + \eta g^2}  {g^{\prime 2} +  g^2}    \cos^2{2\beta}, 
\quad \quad
\eta=\frac{1 + \frac{g_I^2 M_\Sigma^2}{g^2 M_X^2}}
{1+ \frac{ M_\Sigma^2}{ M_X^2}},
\label{mhSU2}
\end{equation}
where
this time $M_\Sigma$ is the soft breaking mass of the scalar in $\Sigma$ and $M_X$ the mass of the quasi-degenerate heavy gauge triplet vectors. Note that both in (\ref{mhU1}) and in (\ref{mhSU2}) the standard MSSM bound is recovered in the supersymmetric limit, $M_\phi, M_\Sigma << M_X$, as it should.
\item {\it $\lambda$SUSY.} This is the NMSSM case with an extra chiral singlet $S$ coupled in the superpotential to 
the usual Higgs doublets by $\Delta f = \lambda S H_1 H_2$, where the upper bound on the lightest scalar is:
\begin{equation}
m_h^2 \leq m_Z^2 (\cos^2{2\beta} + \frac{2\lambda^2}{g^2 + g^{\prime 2}}\sin^2{2\beta}) \, .
\label{mhlsusy}
\end{equation}
\end{itemize}
Mixed cases with extra contributions to the Higgs potential both from  D-terms and from F-terms are also possible, but they are not of interest here since they would not change any of our conclusions. 

\begin{figure}[thb]
\begin{center}
\includegraphics[width=0.55\textwidth]{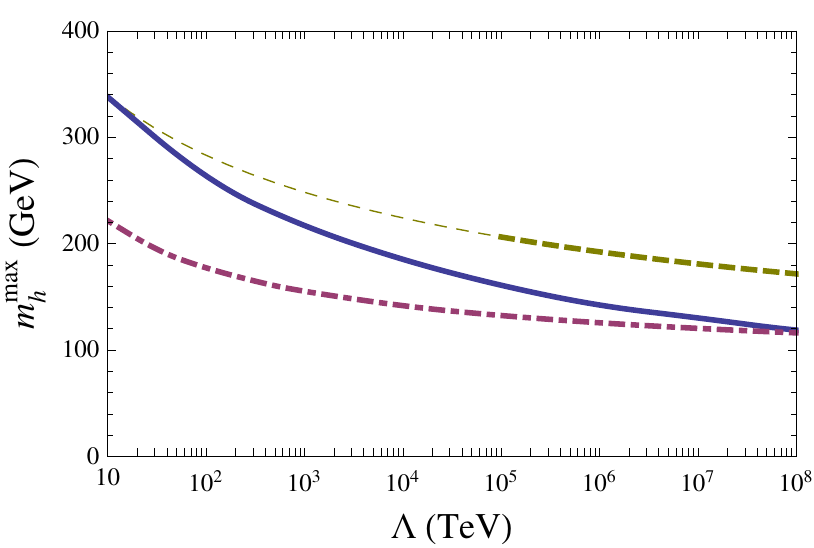}
\end{center}
\caption{{\small  Upper bounds on $m_h$ as function of the scale $\Lambda$ where some coupling starts becoming semi-pertubative, $g_x^2, g_I^2, \lambda^2 = 4\pi$ for the $U(1)$ case (dotdashed, $\tan{\beta} >> 1$), $\lambda$SUSY (solid,  low $\tan{\beta}$) and $SU(2)$ (dashed, $\tan{\beta} >> 1$). In the $SU(2)$ case values of $m_h \gtrsim 200$ GeV are hardly compatible with naturalness and  the EWPT.}}
\label{MaxMhThreeModels}
\end{figure}

Fig. \ref{MaxMhThreeModels} shows the maximal value of $m_h$ in the three different cases ($\tan{\beta} >> 1$ for the extra-gauge cases and low $\tan{\beta}$ for $\lambda$SUSY) as function of the scale at which some coupling becomes semi-perturbative, i.e. $g_x^2 =4\pi$ or $g_I^2 =4\pi$ or $\lambda^2 = 4\pi$. While the bound for $\lambda$SUSY follows straightforwardly 
from (\ref{mhlsusy}) and the renormalization-group running of the coupling $\lambda$, the bounds in the gauge cases include as well the maximal values of $M_{\phi, \Sigma}/M_X$ consistent with naturalness of the heavy scale $M_X$ (10$\%$ fine-tuning at most)\cite{Lodone:2010kt}.
In the $SU(2)$ case values of $m_h \gtrsim 200$ GeV are hardly compatible with the EWPT, coupled with naturalness, due to the large coupling to matter of the extra gauge bosons.

\subsection{Naturalness bounds on the first and second generation s-fermions}

 Having succeeded in raising the Higgs boson mass, we can now ask what happens of the bounds in (\ref{natbounds1}, \ref{natbounds2}). The bound on the stop masses is certainly relaxed, but the value of the stop masses is anyhow no longer relevant to the Higgs mass problem, which is solved by the tree level large extra contributions in all cases. What about the bounds on the sfermion masses of the first two generations? How do they compare with those in Fig. \ref {naturalnessMSSM} for the MSSM?

\begin{figure}[hbt]
\begin{center}
\begin{tabular}{cc}
\includegraphics[width=0.44\textwidth]{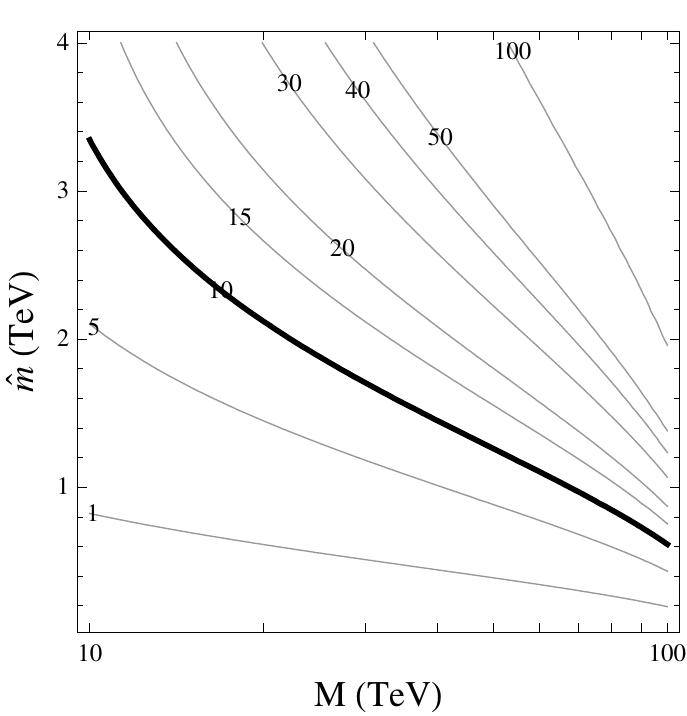} &
\includegraphics[width=0.44\textwidth]{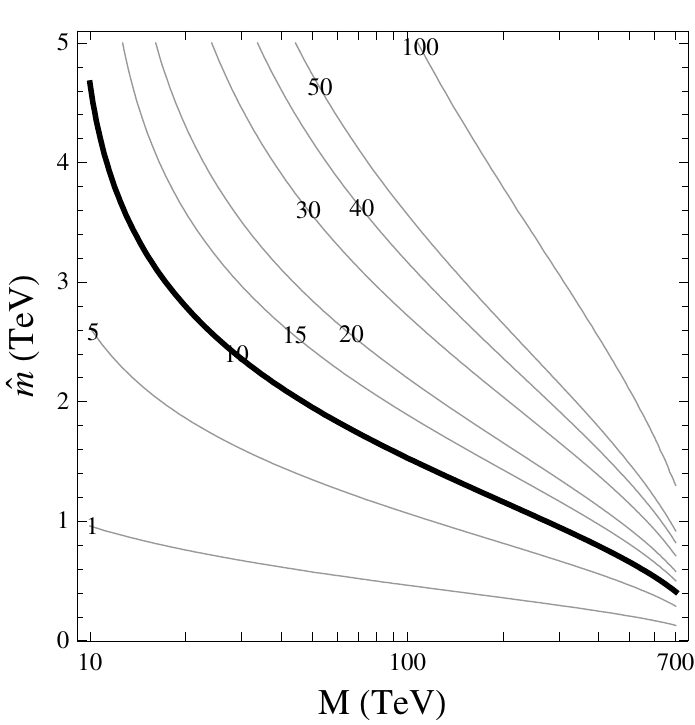}
\end{tabular}
\end{center}
\caption{{\small As in Fig. \ref{naturalnessMSSM} with degenerate scalars at M.
Left: $U(1)$, $m_h = 180$ GeV. Right: $SU(2)$, $m_h = 200$ GeV. }}
\label{naturalnessU1SU2}
\end{figure}

Let us consider the case in which the first two generations of s-fermions take a common value, $\hat{m}$, at a scale $M$, when the dominant effects on the renormalization of $m_h$ come from two loops and the relevant equation in the MSSM case is ($\tan{\beta} >> 1$)
\begin{equation}
\frac{d m_h^2}{d \log{\mu}} = \frac{48}{(16\pi^2)^2}(g^4 + \frac{5}{9} (g^{\prime})^ 4) \hat{m}^2.
\end{equation}
The corresponding equations in the gauge extensions described above are:
\begin{itemize}
\item {\it Extra $U(1)$ factor} 
\begin{equation}
\frac{d m_h^2}{d \log{\mu}} = \frac{48}{(16\pi^2)^2}(g^4 + \frac{5}{9} (g^{\prime})^ 4 + \frac{7}{6}g^4_x)) \hat{m}^2
\label{U1run}
\end{equation}
\item {\it Extra $SU(2)$ factor}
\begin{equation}
\frac{d m_h^2}{d \log{\mu}} = \frac{48}{(16\pi^2)^2}(g_I^4 + \frac{5}{9} (g^{\prime})^ 4) \hat{m}^2
\label{SU2run}
\end{equation}
\end{itemize}
with a clear correspondence between the different equations. From (\ref{natbounds2}), by integrating these equations from $M$ all the way down to $\hat{m}$ itself, one obtains the naturalness bounds shown in Fig. \ref{naturalnessU1SU2}  for fixed values of $m_h$. Note that the running of $\hat{m}$ is by itself negligible since all gauginos are taken significantly lighter. In turn this means that $\hat{m}$ represents a typical mass of any of the s-fermions  of the first two generations, still essentially not split even at $\mu = \hat{m}$.

The comparison of Fig. \ref{naturalnessMSSM} with Fig. \ref{naturalnessU1SU2} makes clear what happens. The presence in (\ref{U1run}) and (\ref{SU2run}) of the contributions from the largish couplings, which are the very source of the increased Higgs boson mass, makes the bound on $\hat{m}$ actually stronger than in the MSSM case. In the $SU(2)$ case  this pattern is insensitive to the way in which the couplings of the matter fields are spread among the two different $SU(2)$ factors, although this may influence the high energy behaviour of the extra gauge couplings themselves.
\begin{figure}[hbt]
\begin{center}
\begin{tabular}{cc}
\includegraphics[width=0.44\textwidth]{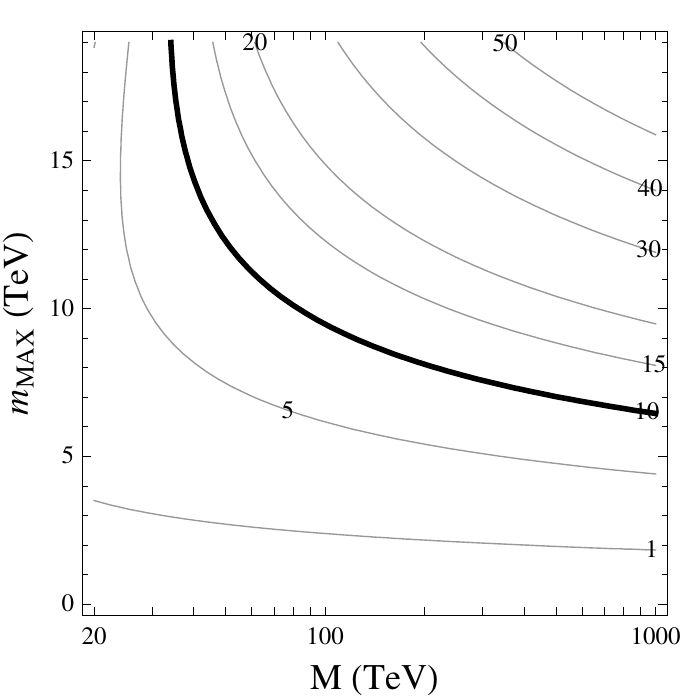} &
\includegraphics[width=0.44\textwidth]{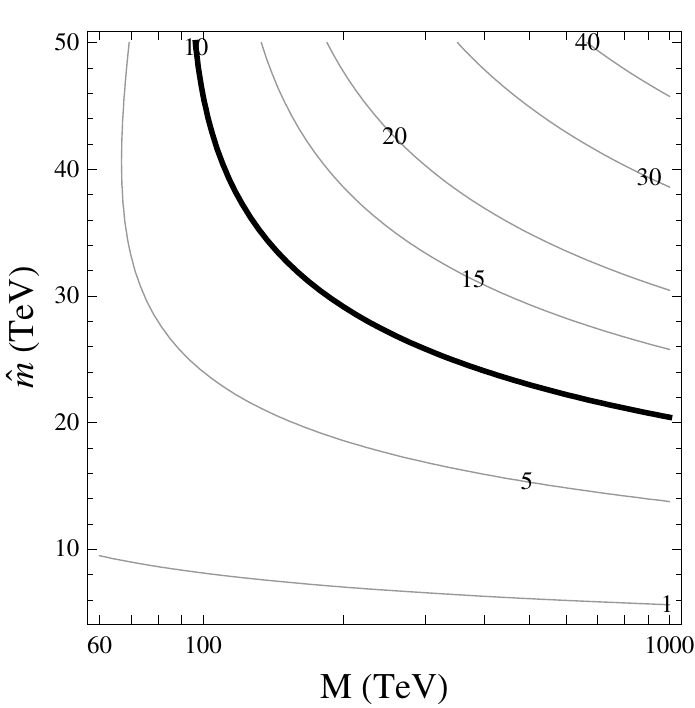}
\end{tabular}
\end{center}
\caption{{\small As in Fig. \ref{naturalnessMSSM} for $\lambda$SUSY, $m_h = 250$ GeV. Left: no special conditions at $M$. Right:  degenerate scalars at $M$. }}
\label{naturalnessLambdaSUSY}
\end{figure}

The situation is completely different in $\lambda$SUSY. Here the Higgs sector is affected by  the largish coupling $\lambda$, but this is essentially not the case for the first two generations of s-fermions due to their negligibly small Yukawa coupling. As a consequence, while the loop dependence of $m_h$ on $\hat{m}$ is the same as in the MSSM, $m_h$ itself is increased, thus reducing the fine tuning. This is shown in Fig. \ref{naturalnessLambdaSUSY} with or without degenerate initial conditions for the s-fermions of the first two generations. For low enough values of $M$, the masses of the first two generations of s-fermions can  go up to $20\div 30$ TeV in a natural way, a factor of $3\div 4$ above  the values in the MSSM.
In view of the considerations developed in Sect. \ref{Hier},  this goes in the direction of solving the supersymmetric flavour problem.

\subsection{Constraint from colour conservation}

As pointed out in \cite{ArkaniHamed:1997ab}, there is an additional constraint on the soft masses of the sfermions of the first two generations. Since colour and electromagnetism must be unbroken, the squared masses of the lighter sfermions of the third generation must not become negative. Neglecting the Yukawa couplings and focussing on the quark sector the relevant RGEs are, up to two loops, with a degenerate initial condition $\hat{m}$ for the first two generations:
\begin{eqnarray}
\frac{d  m_{\tilde{u}_3}^2}{d \log \mu} &=& - \frac{1}{16\pi^2} \, \frac{32}{3} g_3^2 M_g^2 +\frac{8}{(16\pi^2)^2}    \left( \frac{16}{15} g_1^4 + \frac{16}{3} g_3^4 \right)\, \hat{m}^2  \label{u3evolution}
\\
\frac{d  m_{\tilde{Q}_3}^2}{d \log \mu} &=& - \frac{1}{16\pi^2} \, \frac{32}{3} g_3^2 M_g^2 + \frac{8}{(16\pi^2)^2}   \left( \frac{1}{15} g_1^4 + 3 g_2^4 + \frac{16}{3} g_3^4 \right)\, \hat{m}^2   \label{Q3evolution}
\end{eqnarray}
where we also neglected all the gauginos except the gluino.
From (\ref{u3evolution}) and (\ref{Q3evolution}) we see that a large $\hat{m}$ tends to induce negative stop squared masses at the low scale, especially in the case of $\tilde{Q}_3$.

To find a bound on $\hat{m}$ from these considerations we proceed as follows.
First of all we take the value of $m_{\tilde{Q}_{3}}
=m_3$ at $M$ which gives at most 10 \% finetuning on the Fermi scale and comes from:
\begin{equation}
\frac{\partial \log v^2}{\partial \log m_3^2} \approx \frac{6 \, (m_t / \mbox{175 GeV})^2}{16 \pi^2} \, \frac{m_3^2}{m_h^2/2}  \, \log \frac{M}{\mbox{200 GeV}} \leq 10
\end{equation}
which is valid both for the MSSM with large $\tan \beta$ ($m_h=m_Z$) and for $\lambda$SUSY with $\tan\beta\approx 1$ ($m_h = \lambda v$).
Then, starting from this value at the scale $M$, we impose that the running due to (\ref{Q3evolution}) does not drive $m_{\tilde{Q}_3}^2$ negative at $200$ GeV.

\begin{figure}[hbt]
\begin{center}
\begin{tabular}{cc}
\includegraphics[width=0.44\textwidth]{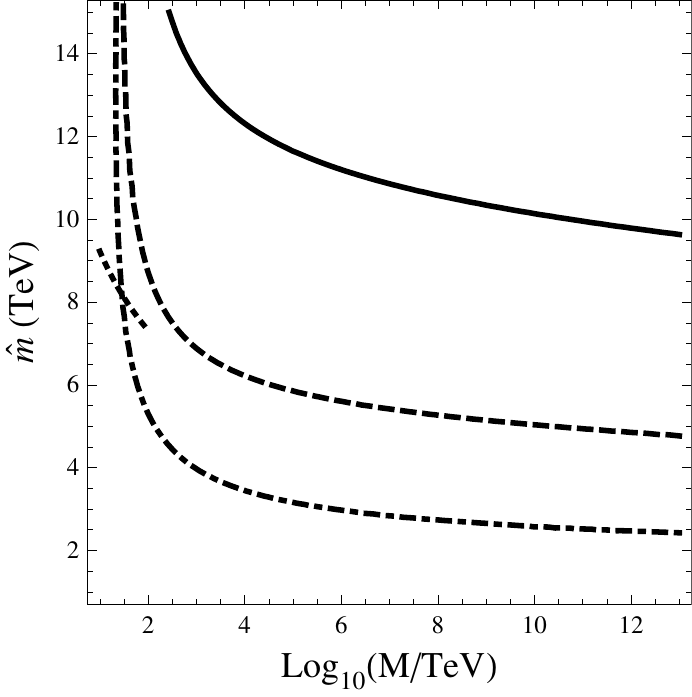} &
\includegraphics[width=0.45\textwidth]{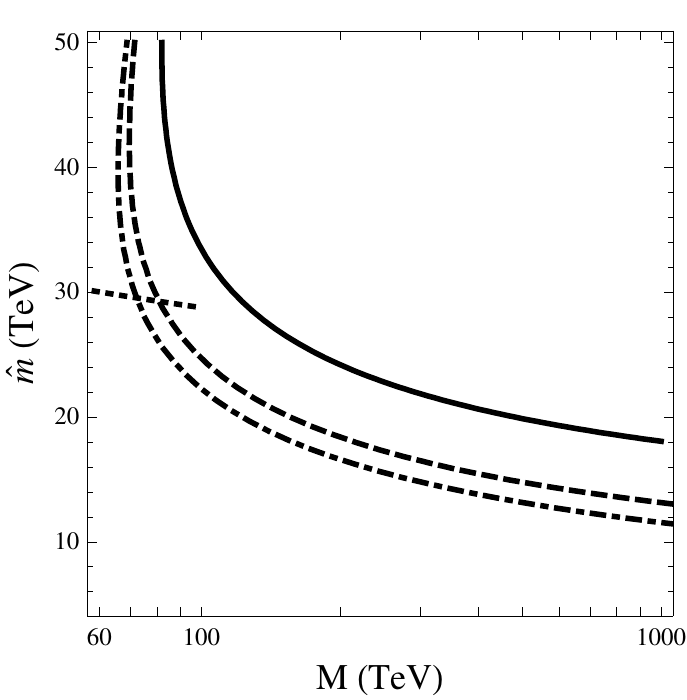}
\end{tabular}
\end{center}
\caption{{\small Colour conservation bound on $\hat{m}$ with ${M}_g=$ 2 TeV (solid), 1 TeV (dashed), 500 GeV (dotdashed). The dotted line below $M=100$ TeV stands for the estimate $\hat{m}/m_{\tilde{Q}_3} \lesssim 25$ for $\hat{m}\sim M$ from \cite{Agashe:1998zz}. Left for $m_h=m_Z$, right for $m_h=250$ GeV. }}
\label{fig:colorunbreaking}
\end{figure}

The result is shown in Figure \ref{fig:colorunbreaking} in the case of the MSSM (left) and in the case of $\lambda$SUSY with $\lambda v$= 250 GeV (right), as a function of $M$, $\hat{m}$, and the gluino mass at low energy ${M}_g$.
Notice that, in the case of the MSSM, for $M=M_{GUT}$ we obtain basically the same bound as in Figure 2 of \cite{ArkaniHamed:1997ab}, with the proper translation of the parameters\footnote{Our colour conservation constraint is actually slightly stronger because we keep only the gluino mass, while \cite{ArkaniHamed:1997ab} keeps all the gauginos with equal mass at $M_{GUT}$.}.
In the case $\hat{m}\sim M$ an important contribution comes from threshold effects, which can be estimated \cite{Agashe:1998zz} to give a bound $\hat{m}/m_{\tilde{Q}_3} \lesssim 25$. This estimate is shown as a dotted line in Figure \ref{fig:colorunbreaking}.

The conclusion is that also this constraint is relaxed in the case of interest, and is not significantly different than the one in Figure \ref{naturalnessLambdaSUSY}.
The relaxation of the bound is due to the fact that we consider a low $M$ scale and moreover, with the same 10 \% finetuning, we can allow stop masses at $M$ which are larger than usual, because of the increased quartic coupling of the Higgs sector. On the contrary, the stronger bounds quoted in the literature \cite{ArkaniHamed:1997ab}\cite{Agashe:1998zz} refer to the case $m_h = m_Z$ and in most cases to $M=M_{GUT}$.

\section{Phenomenological consequences}
\label{pheno}

In this Section we find it useful to outline in an unified way the main phenomenological features of $\lambda$SUSY, leaving a more detailed study to a future work.

\subsection{Gluino pair  production and decays}

At least in a first stage of the LHC and taking into account the current Tevatron constraints, gluino pair production is the source of the relatively most interesting signals.   Naturalness considerations highlight a most crucial region of mass parameters for the gluino, $\tilde{g}$, the two stops, $\tilde{t}_{1,2}$ and for the $\mu$ parameter:
\begin{equation}
m_{\tilde{g}} \lesssim 1800~GeV, \quad\quad
m_{\tilde{t}_1} < m_{\tilde{t}_2} \lesssim 800~GeV, \quad\quad 
\mu \lesssim 400~GeV.
\label{ranges}
\end{equation}
A relevant completion of this set of physical parameters is obtained by adding the mixing angle $\theta_t$
\begin{equation}
\begin{pmatrix}
\tilde t_L\\
\tilde t_R
\end{pmatrix}=\begin{pmatrix}
\sin\theta_t && \cos\theta_t \\
-\cos\theta_t && \sin\theta_t 
\end{pmatrix} \begin{pmatrix}
\tilde t_1\\
\tilde t_2
\end{pmatrix},
\end{equation}
which also determines the mass of the left-handed sbottom, $\tilde{b}_L$\footnote{We neglect the chirality mixing between the two sbottom states,  which is in particular not enhanced by large $\tan{\beta}$ as in the MSSM case. We neglect also small terms in the squark mass-matrices squared proportional to $g^2 v^2$.}, 
\begin{equation}
m_{\tilde b}^2\approx \frac{m_{\tilde t_2}^2-m_{\tilde t_1}^2}{2} \cos 2 \theta_ t+\frac{m_{\tilde t_2}^2+m_{\tilde t_1}^2}{2}-m_t^2,
\end{equation}
the usual gaugino masses $M_{1,2}$ and  the  mass of the right handed sbottom, $\tilde{b}_R$, in the range:
\begin{equation}
\theta_t = 0\div \frac{\pi}{2}, \quad\quad
M_{1,2} \lesssim 600~GeV, \quad\quad 
m_{\tilde{b}_R} \lesssim 600~GeV.
\end{equation}
The upper range for $M_{1,2}$ and $m_{\tilde{b}_R}$ is not relevant to naturalness but has the meaning of a practical decoupling value  for the corresponding particles, given the ranges in (\ref{ranges}). The masses of the third generation sleptons are relatively less important to the phenomenology of gluino decays as long as the Lightest Supersymmetric Particle (LSP) is a neutralino.

An effective way to characterize the signal from gluino pair production is to consider the semi-inclusive Branching Ratios \cite{Barbieri:2009ev}:
\begin{equation}
B_{tt} = BR(\tilde{g}\rightarrow t \bar{t} \chi)\quad\quad
B_{tb} = BR(\tilde{g}\rightarrow t \bar{b} \chi) = BR(\tilde{g}\rightarrow  \bar{t} b \chi)\quad\quad
B_{bb} = BR(\tilde{g}\rightarrow b \bar{b} \chi),
\end{equation}
where $\chi$ stands for the LSP plus $W$ and/or $Z$ bosons, real or virtual, that may occur in the chain decays.
To an excellent approximation in the ranges (\ref{ranges}) it is:
\begin{equation}
B_{tt} + 2 B_{tb} + B_{bb} \approx 1,
\end{equation}
so that the final state from gluino pair production is:
\begin{equation}
pp \rightarrow \tilde{g} \tilde{g}\rightarrow q q\bar{q} \bar{q} +\chi\chi
\label{4q}
\end{equation}
with $q$ either a top or a bottom quark for a total of nine different possibilities.

A particularly interesting signal are the equal-sign di-leptons ($e$ or $\mu$) from semi-leptonic top decays \cite{Barnett:1993ea}\cite{Guchait:1994zk}\cite{Toharia:2005gm}\cite{Acharya:2009gb}, with an inclusive branching ratio:
\begin{equation}
BR(l^\pm l^\pm) = 2 B_l^2 (B_{tb} + B_{tt})^2
\end{equation}
where $B_l = 21\%$. Since $B_{bb}$ is relatively disfavored by $\lambda_t >> \lambda_b$, in the greatest part of the relevant parameter space  $BR(l^\pm l^\pm )$ is between 2  and 4 $\%$. Lower values can occur when: i) 
  $\tilde{b}_{L}$ or   $\tilde{b}_{R}$ become the lightest squarks ( for   $\tilde{b}_{L}$  this is for $\theta_t \rightarrow \pi/2$) and/or ii) $m_{\tilde{g}} \lesssim m_{LSP}+ m_t$. Additional although typically softer leptons can be present in the final states due to W and or Z decays included in $\chi$.

\subsection{A largely unconventional Higgs sector}

The Higgs system of $\lambda$SUSY has been studied in detail in \cite{Barbieri:2006bg}\cite{Cavicchia:2007dp}, although for an almost limit value of $\lambda =2$ and for a relatively heavier singlet scalar $\phi_S$ so that its mixing with the more MSSM-like states, $h, H, A$, can be ignored.

Needless to say a most striking feature of $\lambda$SUSY would be the discovery of the golden mode $h\rightarrow ZZ$, with two real $Z$ bosons, in association with a supersymmetric signal as described above.
The constraint from $b\rightarrow s + \gamma$ is straightforwardly satisfied, given the moderate value of $\tan{\beta}$, for a charged Higgs boson, $H^\pm$, heavier than about 400 GeV, thus implying in most of the parameter space a similar lower bound for the neutral scalars, $H$ and $A$. In turn naturalness suggests all of them not to be heavier than about 800 GeV.

In this  Higgs boson sector, beyond the mass values, there are several important effects due to the largish coupling $\lambda$. One such effect is in the one loop corrections to the $T$-parameter due to the virtual Higgs exchanges. These corrections are positive and automatically of the right size to compensate for the growth of both $T$ and $S$ due to the heavier $m_h$, so as to keep agreement with the EWPT in a relatively broad range of $\tan{\beta}$, not too far from unity \cite{Barbieri:2006bg}. Specifically in the heavy Higgs sector, 
a most striking feature of $\lambda$SUSY is the width for the decay $H\rightarrow hh$, which, being proportional to $\lambda^2$, can go up to  about 20 GeV for $m_H = 500\div 600$ GeV \cite{Cavicchia:2007dp}.

\subsection{Dark Matter: relic abundance and direct detection}

In  $\lambda$SUSY  the LSP can acquire, relative to the MSSM, an extra component in the direction of the neutral singlet $S$. Here we shall consider the case in which such component is negligible, due to its heaviness relative to $\mu, M_1$ and possibly $M_2$. This  allows us to illustrate in clear terms a generic feature of the relic abundance of $\chi_{LSP}$ due to the heaviness, relative to the MSSM, of the lightest Higgs boson. Such feature would in fact be common to any of the models discussed in Sect.  \ref{nolightH} as long as they share a Higgs boson in the $200\div 300$ GeV mass range. 

\begin{figure}
\begin{center}
\begin{tabular}{cc}
\includegraphics[width=0.44\textwidth]{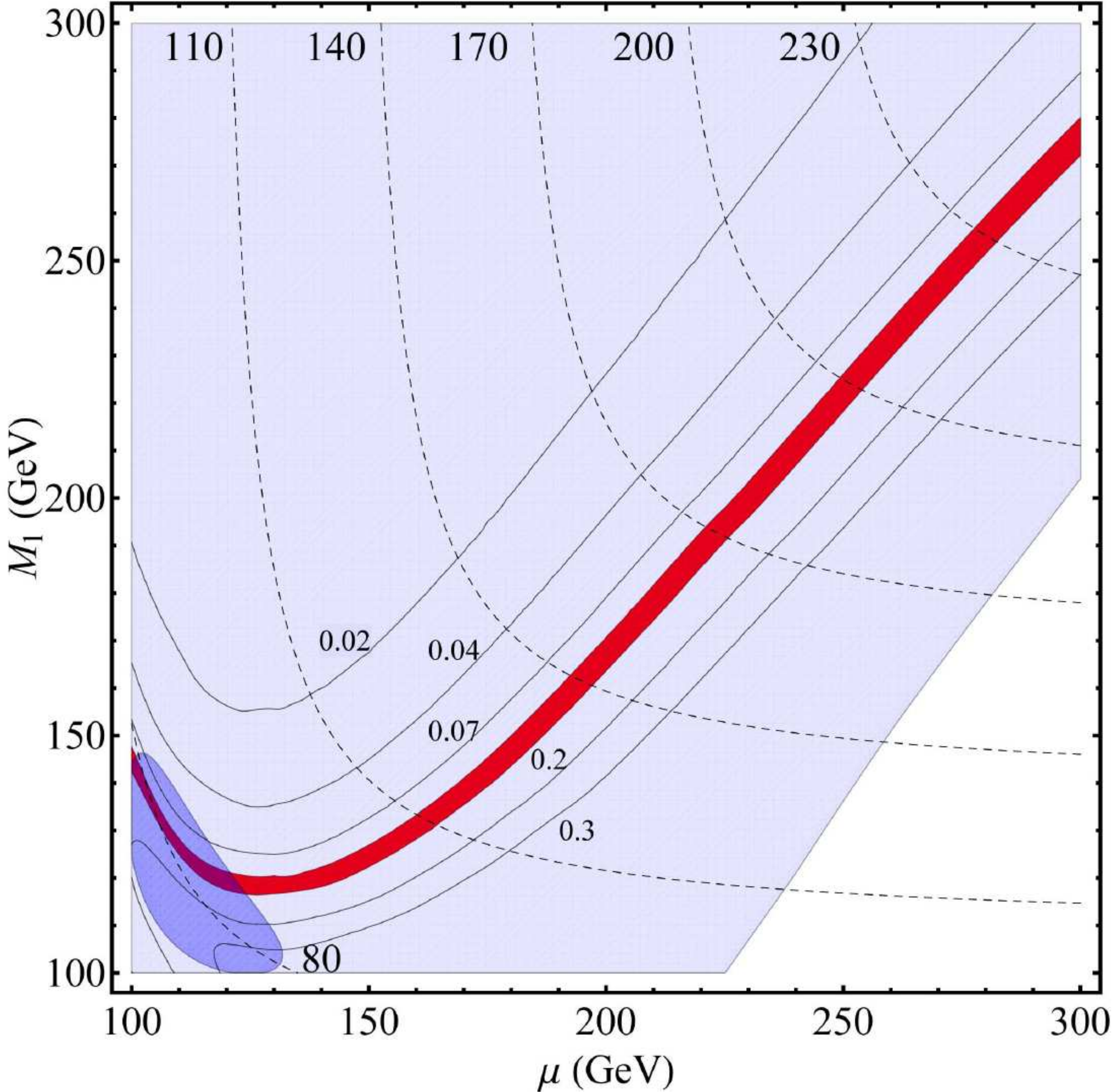} &
\includegraphics[width=0.44\textwidth]{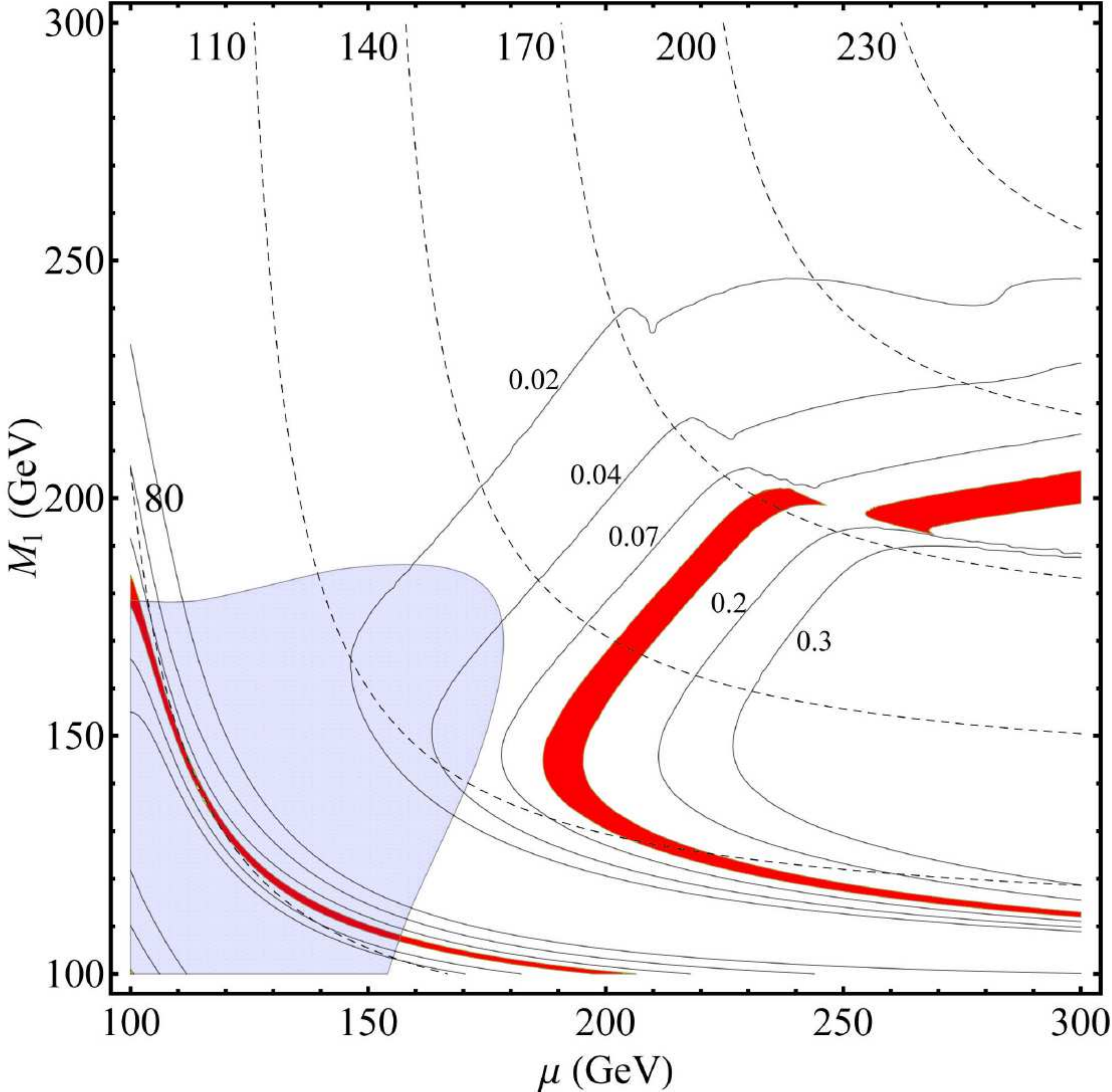}
\end{tabular}
\caption{{\small Isolines of DM relic abundance (solid) and of LSP masses (dashed) for $M_2>> M_1$. Dark blue regions (current CDMS exclusion), light blue (projected XENON100 sensitivity). Left: MSSM, $m_h = 120$ GeV, $\tan{\beta}=7$. Right:  $\lambda$SUSY,  $m_h=200$ GeV, $\tan{\beta}=2$}. }
\label{DM1-2}
\end{center}
\end{figure}

\begin{figure}
\begin{center}
\begin{tabular}{cc}
\includegraphics[width=0.44\textwidth]{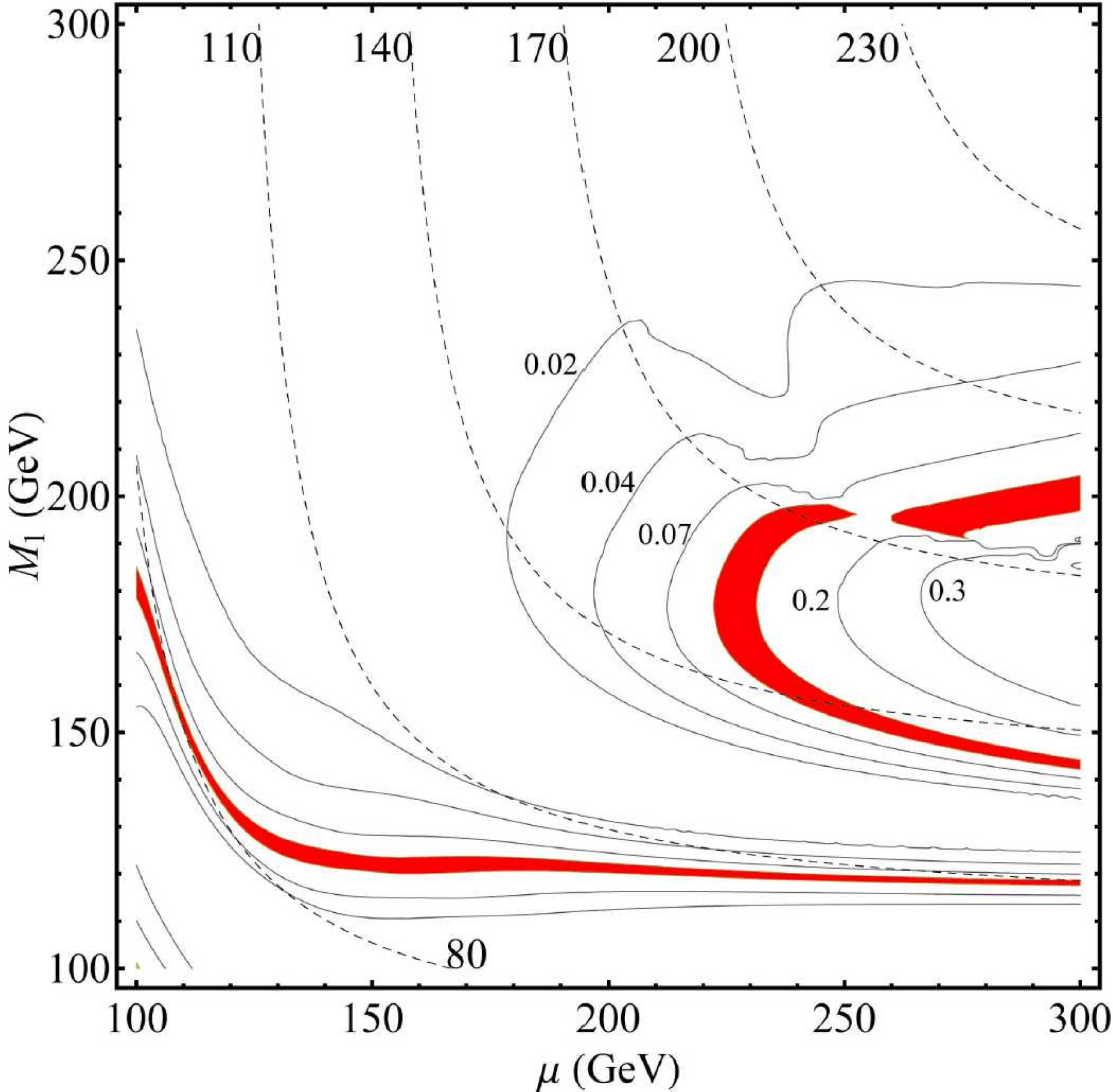} &
\includegraphics[width=0.44\textwidth]{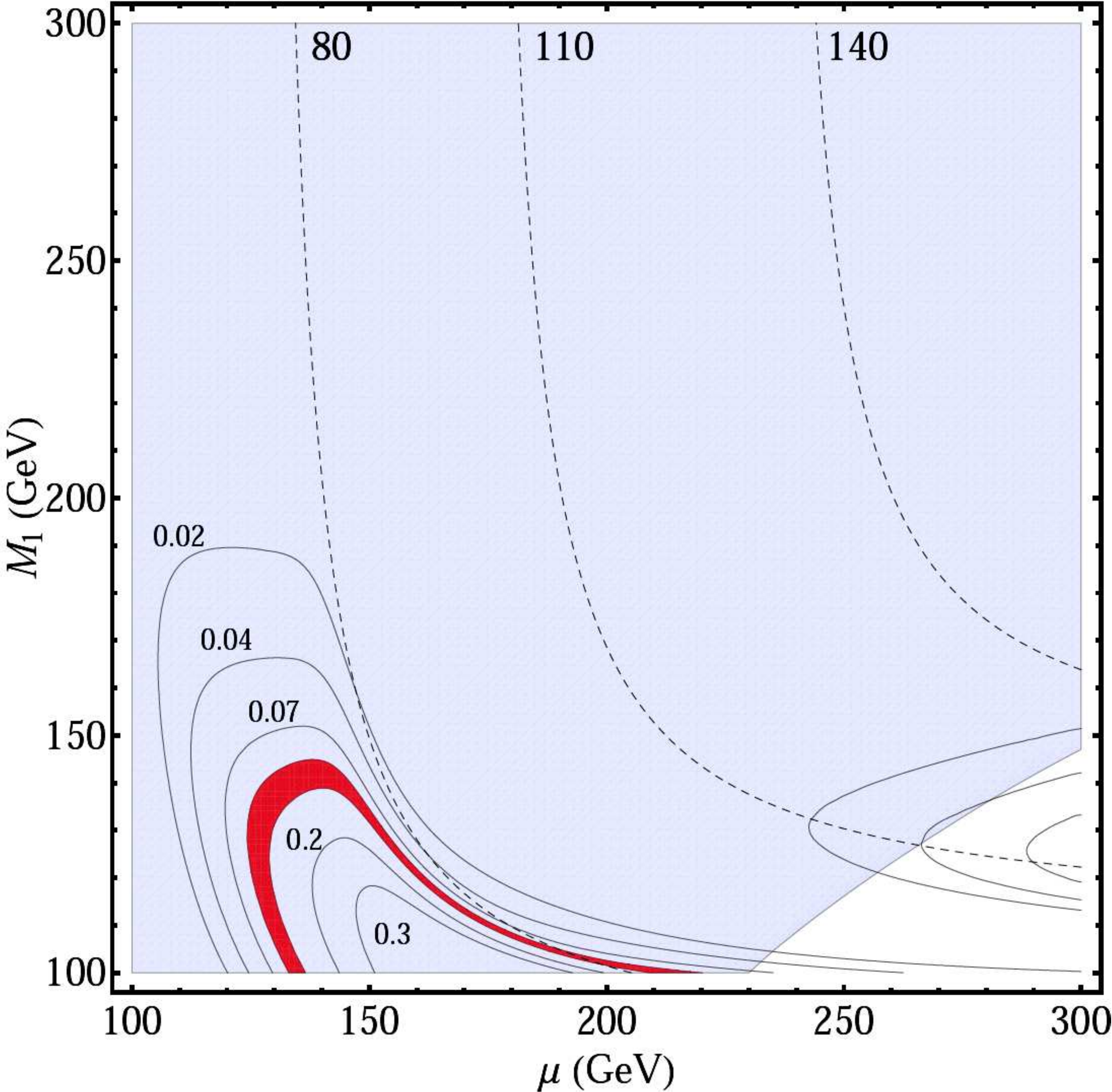}
\end{tabular}
\caption{{\small As in \ref{DM1-2}. 
Left: $\lambda$SUSY, $m_h=250$ GeV, $\tan{\beta}=2$,  $M_2>>M_1$.
Right:  $\lambda$SUSY, $m_h=200$ GeV, $\tan{\beta}=2$, $M_2 = 200$ GeV}.}
\label{DM3-4}
\end{center}
\end{figure}

The way in which the LSP in the MSSM can acquire the observed relic abundance to allow its interpretation as a DM candidate is well known. As  observed in \cite{ArkaniHamed:2006mb}, after LEP constraints are taken into account, the correct prediction for the DM density requires special relations among parameters, justifying  the terminology of ``well-tempered" neutralino. This is neatly illustrated in Fig. \ref{DM1-2} on the left hand side, which is appropriate to the ``well-tempered" bino/higgsino case, i.e. for large (and irrelevant) $M_2$:  to obtain the observed relic abundance, $M_1$ and $\mu$ should be pretty close to each other. In the same plot, which is for $m_h = 120$ GeV and $\tan{\beta} = 7$, the regions are also shown to which the direct detection searches are either currently sensitive \cite{Ahmed:2009zw}  or should become sensitive in a near future \cite{Aprile:2009yh}. To draw these contours we assume everywhere a standard DM density in the halo of our galaxy. These sensitivity regions are therefore directly relevant only where they overlap with regions of correct relic abundance.

The effect of the larger $m_h$ is clearly visible in the same  Fig. \ref{DM1-2} on the right hand side, which is appropriate to $\lambda$SUSY for $m_h = 200$ GeV and $\tan{\beta} = 2$, while $M_2$ is still kept large. In both plots of Fig. 
\ref{DM1-2} it is $m_A=550$ GeV. The effect of the $t \bar{t}$ threshold, only visible in the figure on the right, is due to the $1/\tan{\beta}$ behaviour of the $A t \bar{t}$ coupling,  negligible in the case of the MSSM  for $\tan{\beta}=7$.

Fig. \ref{DM3-4} shows two other cases for $\lambda$SUSY. On the left hand side everything is as in Fig. \ref{DM1-2} right, except for $m_h= 250$ GeV. On the right hand side,  for $m_h= 200$ GeV and  $\tan{\beta} = 2$, $M_2$ is lowered to 200 GeV.
The raise of $m_h$ has also a clear and well known effect on the direct detection cross sections, dominated by $h$-exchange and therefore proportional to $1/m_h^4$ \cite{Barbieri:1988zs}\cite{Drees:1993bu}. This effect is relatively compensated  in the low $M_2$ case by a significant change in the LSP composition.

\section{Conclusions}

Can it be that the Higgs mass problem and the flavour problem contain a unique message? This is the question we have addressed in this work. Notwithstanding the validity of the standard MSSM approach, which makes its test a crucial task of the LHC, 
we believe that this is a meaningful question. Truly enough it rests on the notion of naturalness, which can hardly be viewed as the basis of any theorem. The lack of any serious understanding of the flavour pattern is another difficulty we face.
Yet the possibility that the Higgs mass problem and the flavour problem point to an extension of the MSSM needs to be given serious consideration. The basic simple idea that we pursue is that a lightest Higgs boson naturally heavier than in the MSSM renders at the same time more plausible that the supersymmetric flavour problem has something to do with a hierarchical structure of the s-fermion masses, a connection often invoked in the past. 

At first the constraints set by the lack of flavour signals would seem to require values of the masses of the first two generations totally incompatible with naturalness. However the combination of mild flavour assumptions with a relaxation of the naturalness constraints by an order of magnitude thorough a heavier Higgs boson than normal can change the situation. 
The concrete proposal that we make, which should at least be taken as an example, consists of the following. With degeneracy and alignment between the first two generations of s-fermions controlled by a parameter of the order of the Cabibbo angle and a ratio of $4\div 5$ between $\delta^{LL}_{12}$ and $\delta^{RR}_{12}$, in one direction or another, even the hardest flavour constraints can be satisfied by $m_{\tilde{f}_{1,2}} \gtrsim 20\div 30$ TeV. 
In turn these masses are natural if they are born degenerate, at least within $SU(5)$ multiplets, at a scale $M$ below $10^3$ TeV and a modification of the Higgs sector, which remains perturbative  up to the same  scale, raises the lighest Higgs boson mass in the $200\div 300$ GeV range, e.g. like in $\lambda$SUSY. No matter what produces it, a Non Standard Supersymmetric Spectrum like the one shown in Fig. \ref{spettro2} is brought into focus.

Even before any detailed investigation, which we believe is well worth doing, the following phenomenological consequences clearly emerge:
\begin{itemize}
\item The abundance of top, even generally more than bottom quarks, in the gluino decays, giving rise to a distinctive signature in gluino pair production, which could be detected already in the early stages of the LHC.
\item The appearance of the very much non MSSM-like golden mode decay of the lighest Higgs boson, $h\rightarrow ZZ$, although with a reduced Branching Ratio \cite{Barbieri:2006bg}\cite{Cavicchia:2007dp} relative to the SM one with the same Higgs boson mass.
\item A distinctive distortion of the relic abundance of the lightest neutralino, again relative to the MSSM, due to the $s$-channel exchange of the heavier Higgs boson in the LSP annihilation cross section, with an LSP which needs no longer be ``well-tempered".
\end{itemize}
Several other phenomenological features, more or less tied to $\lambda$SUSY, are present, which may be useful to study carefully.

\section*{{Acknowledgments}}

{We thank Andrea Romanino and Marco Nardecchia for help on Sect. \ref{Hier}, as well as Gino Isidori and Gilad Perez for their comments on related issues. 
This work is supported in part by the European Programme ``Unification in the LHC Era",  contract PITN-GA-2009-237920 (UNILHC), by the
MIUR under contract 2006022501 and by the Swiss National Science Foundation under contract No. 200021-116372. }



\end{document}